\documentstyle[emulateapj,apjfonts,psfig]{article}

\begin{document}

\newcommand{\be}{\begin{equation}}
\newcommand{\ba}{\begin{eqnarray}}
\newcommand{\ee}{\end{equation}}
\newcommand{\ea}{\end{eqnarray}}
\newcommand{\ceff}{C_{\rm eff}}
\newcommand{\bout}{B_{\rm out}}

\title{Feedback Processes in Early-Type Galaxies}
\author{Ignacio Ferreras$^1$, Evan Scannapieco$^{2,3}$ 
\& Joseph Silk$^1$}
\affil{$1$ Nuclear \& Astrophysics Lab. Keble Road, Oxford OX1 3RH, 
England, UK\\
$2$ Department of Astronomy, University of California, Berkeley\\
$3$ Osservatorio Astrofisico di Arcetri, 50125 Firenze, Italy}
\authoremail{ferreras@astro.ox.ac.uk, evan@arcetri.astro.it
silk@astro.ox.ac.uk}
\slugcomment{Accepted July 11, 2002}

\begin{abstract}
We present a simple phenomenological model of feedback in early-type
galaxies that tracks the evolution of the interstellar medium gas
mass, metallicity, and temperature.  Modeling the star formation rate
as a Schmidt law with a temperature-dependent efficiency, we find that
intermittent episodes of star formation are common in moderate-size
ellipticals.  Our model is applicable in the case in which the
thermalization time from SN is sufficiently long that spatial
variations are relatively unimportant, an appropriate assumption for
the empirical parameters adopted here, but one that can only be
demonstrated conclusively though more detailed numerical studies.  
The departure from a standard scenario of passive evolution implies
significantly younger luminosity-weighted ages for the stellar
populations of low-mass galaxies at moderate redshifts, even though
the more physically meaningful mass-weighted ages are changed only
slightly.  Secondary bursts of star formation also lead to a natural
explanation of the large scatter in the NUV-optical relation observed
in clusters at moderate redshift and account for the population of E+A
galaxies that display a spheroidal morphology.  As the late-time
formation of stars in our model is due to the gradual cooling of the
interstellar medium, which is heated to temperatures $\sim$ 1 keV by
the initial burst of supernovae, our conclusions do not rely on any
environmental effects or external mechanisms.  Furthermore, a simple
estimate of the X-ray emission from this supernova heated gas leads to
an $L_X$ vs $L_B$ correlation that is in good agreement with observed
values.  Thus feedback processes may be essential to understanding the
observed properties of early-type galaxies from the optical to the
X-ray.

\end{abstract}
\keywords{galaxies: elliptical and lenticular, cD - galaxies: evolution - 
galaxies: stellar content}


\section{Introduction}

Of all the fields of astronomy, the study of galaxy formation may be
the one in which the rift between theory and observation is the most
difficult to bridge.  Theoretical astrophysicists, perhaps from a
temperamental as well as a practical point of view, are best at
predicting the most difficult component of the universe to measure:
the overall distribution of invisible ``dark matter.'' While
the evolution of this component is well understood,
numerical and analytical models must stretch their predictive powers
to the limit to superimpose on this history the messy heating,
cooling, and enrichment processes that affect baryonic gas.

At the same time, observational astrophysicists, perhaps from a
romantic as well as practical point of view, are happiest when looking
at the most complicated thing they could ever measure: the stars in
the night sky.  A proverbial tail that wags the dog, the magnitudes
and colors of the stellar populations in galaxies are dependent not
only on the history of the baryonic gas, but on the notoriously
complicated process of star-formation.  As many possible
environmental effects can have an unknown impact on the number and
distribution of stars formed (See, eg. Kroupa 2001; Larson 1999; Scalo
1998) which is even more uncertain under primordial conditions (see,
eg. Nakamura \& Umemura 2001), predicting the optical properties of
galaxies involved from ``first principles'' requires an enormous
amount of extrapolation and simplifying assumptions.

This great rift calls for astronomers to work to meet each other from
each side of the divide.  Thus simulations and semi-analytical models
of galaxy formation are continuously striving to include all the
important physical processes that affect forming galaxies and the
surrounding intergalactic medium (e.g.  Kauffmann et al.\ 1999; Baugh
et al.\ 1998; Somerville \& Primack 1999).  From the observational
point of view, astronomers must constantly develop
more careful phenomenological models, which better express the
features seen in simulations: making statements as to the
ages of stars, their metallicities, and levels of dust extinction,
rather than simply considering their infrared, optical, and
ultraviolet colors and magnitudes.  Models of this sort include Tinsley
(1980), Matteucci \& Tornamb\'e (1987), Tantalo et al.\ (1998), and
Ferreras \& Silk (2000b). 

From a theoretical point of view one of the most important issues to
recently come to the fore is the role of supernova feedback in galaxy
formation.  N-body simulations have shown that the baryonic components
of galaxies that form in Cold Dark Matter (CDM) simulations
systematically lose too much of their angular momentum to the dark
matter halos in which they are contained (Navarro \& Benz 1991).  As
discussed in White (1994), the resolution of this problem is widely
believed to be the inclusion of feedback, although the proper modeling
of this process remains a subject of intense debate.  Thus since the
first numerical model of thermal heating by Katz (1992), approaches
have ranged from implementation of kinetic boosts (Milos \& Hernquist
1994; Gerritsen 1997), a multiphase model of star formation and
feedback (Yepes 1997; Hultman \& Pharasyn 1999), a model of turbulent
pressure support from feedback (Springel 2000), and a model in which
energy persists in the interstellar medium for a period corresponding
to the lifetime of stellar associations (Thacker \& Couchman 2001).
Finally, feedback processes are likely to play a more general role in
impacting the intergalactic medium and conditions under which galaxy
formation occurs (Mac~Low \& Ferrara 1999; Martel \& Shapiro 2001;
Scannapieco \& Broadhurst 2001; Scannapieco, Thacker, \& Davis 2001).

Yet despite the many exploratory feedback models being examined
theoretically, the role of feedback in empirical modeling has been more
limited.  Ferreras \& Silk (2000a; 2001) explored a simple model in
which feedback was included phenomenologically using two free
parameters: the star formation efficiency and the fraction of gas
ejected in outflows. These values are dependent on the thermal and
mechanical feedback from supernovae but they fail to capture their
impact the state of the interstellar medium (ISM) itself, despite the
close relationship between this state and star formation (McKee \&
Ostriker 1977).  Our primarily theoretical motivation in this paper,
then, is to extend these models by tracking the temperature
evolution of the ISM, taking into account the thermal contributions
of supernovae and metallicity-dependent gas cooling.

The primarily observational motivation for this paper, on the other
hand, is the measurement of a large scatter at the faint end of the
near-ultraviolet (NUV) minus optical color-magnitude relation in
cluster Abell~851 ($z=0.4$) observed by Ferreras \& Silk (2000a) using
passbands F300W and F702W of the WFPC2 on board the Hubble Space
Telescope.  A simple analysis of this scatter shows that it can not be
accounted for by old Horizontal Branch stars.  Moreover, significantly
younger luminosity-weighted ages were found, although a degeneracy
between the age of the young stellar component and its mass fraction
prevented a detailed estimate of a more physically meaningful
mass-weighted age. This result shows that the star formation history
of early-type galaxies is much more complicated than the standard
picture in which a single stellar population formed at a redshift
$z\gtrsim 3$ and then evolved passively.

The structure of this work is as follows.  In \S2 we describe a simple
thermal model of supernova feedback in elliptical galaxies and its
impact on the gas, stellar, and metallicity evolution of these
objects.  In \S3 we describe the general features of our model, and
the physics on which they depend.  Our evolutionary tracks are
combined with population-synthesis models in \S4 and used to study the
color-magnitude relationship in ellipticals in both the optical and
near-ultraviolet.  In \S5 we examine the compatibility of our model
with X-ray observations of early-type galaxies.  In \S6 we discuss
the implications of our modeling for future observational studies of
feedback in ellipticals as well as how our model compares to other
theoretical approaches, and conclusions are given in \S7.

\section{Thermal Model of Feedback in Elliptical Galaxies}

\subsection{Basic Assumptions}

In order to reconstruct the star-formation and chemical enrichment of
early-type galaxies we follow an approach similar to that adopted in
Ferreras \& Silk (2000b) which in turn is an adaptation of the
formalism of Tinsley (1980) in which the evolution 
within a galaxy is reduced to a few general parameters.  Our
model consists of only gas and stars.  For each
component we trace the net metallicity, counting all elements heavier
than helium in the same way.  The mass in gas -- $M_g(t)$ -- and in
stars -- $M_s(t)$ -- in each model are normalized to the initial total
gas mass $M_{g0}$ such that $\mu_g(t)\equiv M_g(t)/M_{g0}$ and
$\mu_s(t)\equiv M_s(t)/M_{g0}$. 

The gas component is fueled by infall of pre-enriched gas at a rate
assumed to be a Gaussian function: 
\be
f(t)\propto\frac{1}{\tau_f\sqrt{2\pi}}\exp
\left[ -\frac{(t-t_{f0})^2}{2\tau_f^2}\right] , 
\ee 
where $\tau_f$ is the
infall timescale and $t_{f0}$ is the epoch of maximum infall,
parameterized in terms of a formation redshift $z_F$ such that $t(z_F)
= t_{f0}$. We additionally assume a Salpeter (1955) Initial Mass Function
(IMF) with cutoffs at $0.1$ and $100 M_\odot$, and an initial
metallicity of the infall gas of $Z_0 = Z_\odot /10$.  This value is
motivated by Renzini (1999) who used an estimated $Z_\odot /3$
metallicity of the $z=0$ Universe and the fact that $\sim 30$\% of all
stars having formed at $z\gtrsim 3$, to imply that the average
metallicity of the high-redshift Universe is $\sim 1/3\times 1/3\sim
1/10$ solar.  Similar levels of pre-enrichment are consistent with the
observed lack of G-dwarf stars with metallicities below $\sim 0.1 Z_\odot$
in the Milky Way disk, and the paucity of such stars in other massive
early and late-type galaxies (Worthey, Dorman, \& Jones 1996; Thomas,
Greggio, \& Bender 1999).


\subsection{Gas and Stellar Evolution} 
We consider an infall model in which the star formation efficiency is
determined by the temperatures of the gas.  In this case the limiting
factor in star formation is not the conversion of cool gas into stars,
but rather the cooling of the gas itself into molecular clouds. We
write the star formation rate as a Schmidt law with respect to the
normalized gas mass:
\be
\psi(t) = \ceff (T) \mu_g^{\cal N}(t),
\ee
where $\ceff (T)$ is the temperature-dependent star formation 
efficiency and $1< {\cal N}<2$. 
A comparison of the Schmidt law with the observed star formation rate
in local galaxies gives a value ${\cal N}\sim 1.5$ (Kennicutt 1998), 
adopted throughout this paper. We assume a Fermi-step function 
for the dependence of the star formation efficiency on 
the temperature:
\be
\ceff (T_4) = \frac{\ceff^0}{1+\exp [(T_4-1)/\alpha ]},
\ee
where $T_4\equiv T/10^4K$ and $\alpha$ is a parameter that controls
the steepness of the step.  This function assumes that the star
formation efficiency decreases abruptly for an average temperature of
the ISM above $T_4\sim 1$, chosen to correspond to the temperature at
which hydrogen is ionized.  Finally, we assume fixed values for the
normalization ($\ceff^0=100$), and the steepness of the step ($\alpha
=3$).

With these definitions the equations for the evolution of gas and
stars can be written as a sum of gas accretion, star-formation,
and ejection of material from stars:
\ba
\frac{d \mu_g}{dt} = f(t) -\ceff(T_4)\mu_g^{\cal N} + (1-\bout)E(t),
\label{eq:mug}\\
\frac{d \mu_s}{dt} =  \ceff(T_4)\mu_g^{\cal N} - E(t),
\ea
where the integral $E(t)$ is the mass of gas ejected at a time $t$
from stars at the end of their lifetimes:
\be
E(t) = \int_{m_t}^\infty dm\phi (m)(m-w_m)
\psi(t-\tau_m-\tau_{\rm sn}),
\ee
where $\phi (m)$ is the initial mass function, $w_m$ is the mass of
a stellar remnant with an initial mass $m$, and $m_t$ is the mass
corresponding to a stellar lifetime $t$, each of which can be fixed
using the values as in Ferreras \& Silk (2000b).  A delay
$\tau_{\rm sn}$ is included to account for the fact that it takes some
time for the energy from supernovae to spread throughout the
galaxy. This should be roughly equal the time it takes for a gas
bubble heated by the surrounding supernovae ejecta to percolate
throughout the ISM of the galaxy, which we estimate to be $\tau_{\rm
sn}\sim 100$~Myr.  Finally, $\bout$ is the fraction of gas ejected in
outflows, as discussed in Ferreras \& Silk (2000b).  This parameter is
a decreasing function of galaxy mass that lies in the range
$0\leq\bout\leq 1$ and determines the average metallicity of the
stellar populations.  This inefficiency of metal ejection in large
ellipticals is thought to be the best explanation of the correlation
between mass and metallicity (Arimoto \& Yoshii 1987) which gives rise
to the color-magnitude relation.

\subsection{Thermal Feedback}

In order to model the temperature evolution of the gas we consider the
cooling and heating sources that contribute to its total energy
content.  Defining ${\cal E}$ as the total
internal energy, we find:
\be
\frac{d{\cal E}}{dt}= - N_e \, n_e \,\Lambda(T,Z) +
	\frac{f}{M_\odot} \, k T_f +
	\epsilon \, 10^{51} {\rm erg} \, \frac{M_{g0}}{M_\odot}
	R_{\rm sn}
\ee 
where $N_e$ and $n_e$ are the total number and number density of
electrons, and $T_f$ is the temperature of the infalling gas, and
$\epsilon$ is the fraction of the SN energy which goes toward
heating the gas.  The three terms on the right-hand side of this
equation correspond to radiative cooling in the gas, thermal input
from infalling gas, and energy input from SNe, respectively.

The first of these is proportional to the electron density of the gas
squared times the radiative cooling function, which is dependent on
both the temperature and overall metallicity of the gas. In our simple
model, we assume a polytropic equation so that $n_e\propto
T^{1/(\gamma -1)}$. The value of the polytropic index $\gamma$ depends
on the state of the ISM as shown below.  The cooling function, on the
other hand, can be calculated directly according to the tabulated
model in Sutherland \& Dopita (1993), which assumes solar abundance
ratios.

In the second term, we take $T_f$ to be the virial temperature of the
halo; and in the last term, $R_{\rm sn}$ is the rate of
supernovae in the galaxy and $\epsilon$ is the fraction of the
resulting mechanical energy that is deposited into the gas.  In
calculating $R_{\rm sn}$ we consider contributions from both
Type II and Type~Ia SNe, such that $R_{\rm sn}(t) = R_{II}(t) +
R_{Ia}(t)$.  The first of these contributions can be written as 
\be
R_{II}(t)=\int_{8M_\odot}^{100M_\odot} dm\phi (m)
\ceff[T_4(t-\tau_m\tau_{\rm sn})] \mu_g^{\cal N}(t-\tau_m-\tau_{\rm sn}), 
\ee
where the integral gives the contribution from massive stars that
undergo core collapse.

In the Type~Ia case, we follow the standard prescription of Greggio \&
Renzini (1983), recently explored in more detail by Matteucci \&
Recchi (2001). In this model the rate is written with respect to the
distribution function of binary systems that can harbor a white
dwarf:
\be
R_{Ia}(t) = {\cal A}\int_{m_t}^{16M_\odot}dm \phi(m)
\int_{\mu(m)}^{\mu_{max}}d\mu^\prime f(\mu^\prime )\psi(t-\tau_m-\tau_{\rm sn}),
\label{eq:Ia}
\ee
where $m_t={\rm max}(3M_\odot ,m(t))$ and $\mu$ is the ratio of the mass
of the secondary star and the total, which ranges in the integral from
$\mu(m) = {\rm max}(1-8/m, 0.8/m)$ to $\mu_{max}=0.5$.  Finally,
$f(\mu ) = 2^{1+\beta}(1+\beta )\mu^\beta$ is the distribution
function of binaries which is taken from Greggio \& Renzini (1983),
with $\beta =2$.  The normalization in this case is taken to be
${\cal A} = 0.05$, calibrated from the ratio of Type~Ia to Type~II
supernovae that best fits solar abundances $N(Ia)/N(II)=0.12$
(Nomoto, Iwamoto, \& Kishimoto 1997).

In order to compute the evolution of the temperature, we assume an ideal 
gas ($p=nkT$), with an equation of state given by $p\propto n^\gamma$,
which implies $T\propto n^{\gamma -1}$. Writing
\be
d{\cal E} = \frac{3}{2}NkdT-pdV,
\ee
we find
\be
 \frac{d{\cal E}}{kN} = dT \left( \frac{3}{2} + \frac{1}{\gamma -1}\right).
\ee
During the formation history of the galaxy we assume a single equation
of state, namely, that of an ideal gas at constant pressure. Hence we
fix $\gamma =0$ throughout. This is motivated by cooling timescales
which are much longer than the dynamical time.
Rewriting the feedback efficiency in terms of the fraction of gas
ejected in outflows, and solving for the temperature in terms of $T_4$,
the evolution of the gas temperature becomes
\ba
2\mu_g\frac{d T_4}{dt} =& (1-\bout) m_{0.7} T_{\rm sn}(R_{II}+R_{Ia})-
[\psi -(1-\bout )E] T_4 - \nonumber \\
& - \frac{T_4}{\tau_{\rm cool}} \frac{N_e\mu_g }{N_g} + f(T_f-T_4),
\ea
where $m_{0.7}$ is the mean mass of a gas particle in units
of $0.7$ times the proton mass (note that this gas is ionized),
$\tau_{\rm cool}\equiv kT/n\Lambda$ is the cooling timescale,
and $T_{\rm sn}\sim 3\times 10^5$ (in units $10^4$~K) is 
again the energy input from supernovae, now rescaled in terms of
a supernova ``temperature,'' which corresponds to a thermal
efficiency of $\epsilon = 0.7 (1-B_{\rm out}).$

Note that our model ignores the thermal contribution from the ejecta
of intermediate and low-mass stars.  This approximation is based on
the assumption that the temperatures of planetary nebulae are much
less than the typical temperature of gas in the galaxy.

\subsection{Chemical Enrichment}

The final element of our model is the evolution of metals 
in the gas.  In this case the relevant equation is
\be
\frac{d(Z_g\mu_g)}{dt} = Z_f f - Z_g \ceff(T) \mu_g^{\cal N} 
	+ (1-1.2\bout) (E_Z+y_{Ia}R_{Ia}), 
\label{eq:zg}
\ee
where $Z_f=0.1 Z_\odot$ is the metallicity of the pre-enriched
infalling gas.  Note that each of these terms parallels those in 
eq.\ (\ref{eq:mug}).  In eq.\ (\ref{eq:zg}) $E_Z$ is the amount of metals 
ejected from stars:
\be
\begin{array}{ccl}
E_Z(t) & \equiv & \int_{m_t}^\infty dm\phi (m)
\psi (t-\tau_m-\tau_{\rm sn})\times\\
 & & \times\big[ (m-w_m-mp_m)Z_g(t-\tau_m-\tau_{\rm sn})+mp_m\big],\\
\end{array}
\ee
where he fraction of a star of mass $m$ transformed into metals is given by
$p_m$. For Intermediate Mass Stars (IMS), with $M \leq 8 M_\odot$
we use a power law fit to the yields from Renzini \& Voli (1981),
Marigo, Bressan \& Chiosi (1996) and van~den~Hoek \& Groenewegen
(1997). For larger stars, which contribute to metal enrichment via
Type-II supernovae, the yields are taken from Woosley \& Weaver (1995)
and Thielemann, Nomoto \& Hashimoto (1996).  Note that the differences
in the yields between these two groups are due to the different
physical inputs assumed: mainly the criterion for convection, the
nuclear reaction rates, and the determination of the upper mass cut.
Finally, the $(1-1.2\bout)$ factor in eq.\ (\ref{eq:zg}) accounts
for the fact that out-


\centerline{\null}
\vskip+3.3truein
\includegraphics{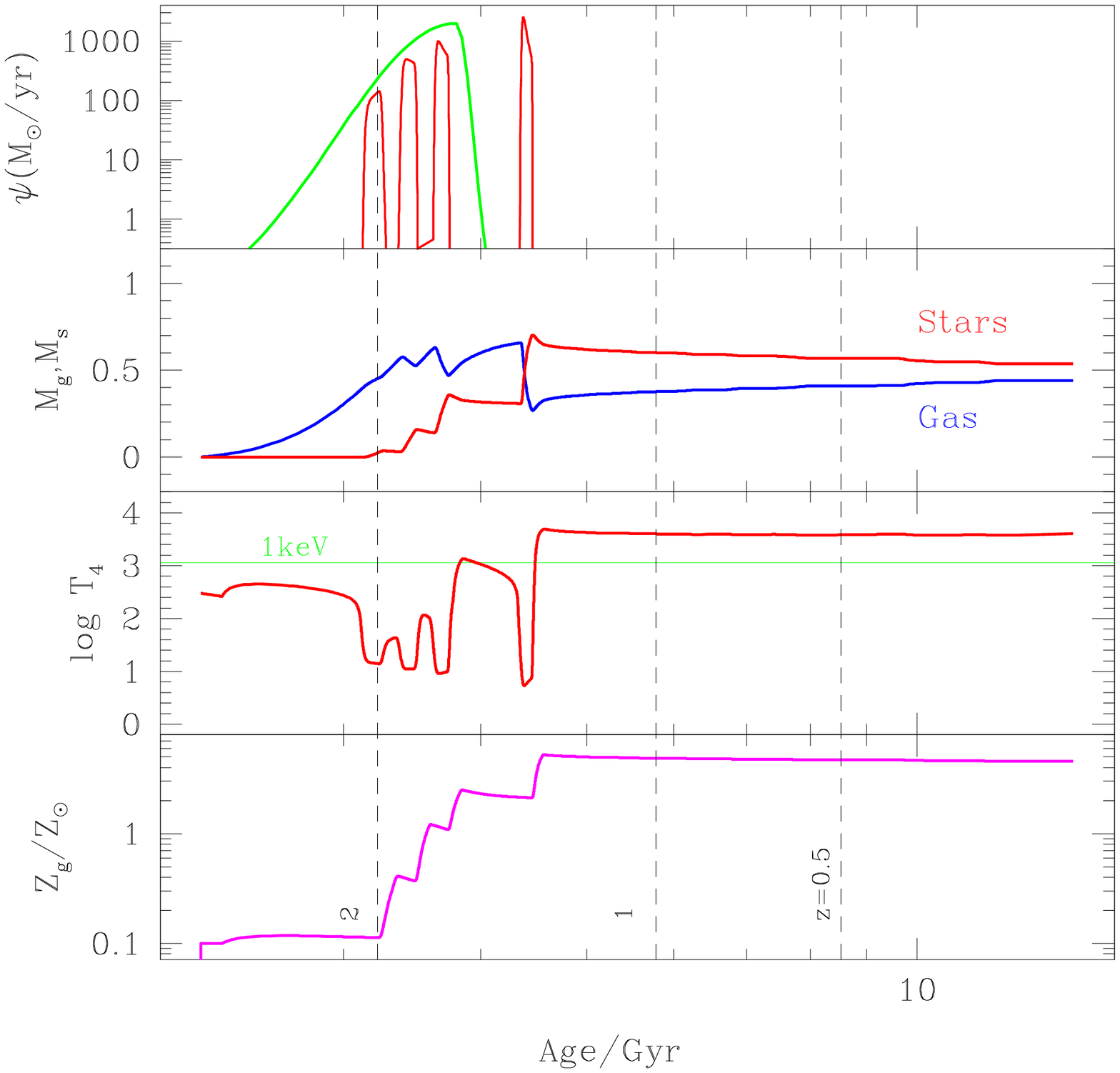}
\figcaption[th1.eps]{
Star formation history of a massive elliptical galaxy, which 
corresponds to $T_0=300$; $n_0=2$; $\tau_f=0.5$~Gyr; $z_F=2$; $\bout=0$. 
The photometry at zero redshift $U-V=1.55$; $V-K=3.25$ corresponds to 
a bright elliptical ($M_V=-22$; $\log\sigma =2.4$) in a local cluster.
\label{fig:big_ellip}}
\vskip+0.2truein


\noindent flows driven by supernova-induced winds 
are likely to be enriched with metals, as described by Vader (1986).
Finally, Type~Ia supernovae must be added to the chemical enrichment
process.  Using the rate of supernovae -- $R_{Ia}(t)$ -- as given by
eq.\ (\ref{eq:Ia}), we include the contribution to the metallicity of
the gas, assuming the yield from each type~Ia supernova to be $y_{Ia}
= 0.6 M_\odot$ (mostly iron; Thielemann, Nomoto \& Yokoi 1986). There
is no such term in the equation tracing the evolution of gas, eq.\
(\ref{eq:mug}), however, since this type of supernova does not
contribute to the gas reservoir.  The standard progenitor of a type~Ia
supernova is a binary system with a C-O white dwarf. Given that the
maximum initial stellar mass that leads to a white dwarf is
approximately $8M_\odot$, this means type~Ia supernova begin to 
contribute $0.03$~Gyr after stars are first formed.


\subsection{The Formation Epoch of Ellipticals}

In keeping with our desire to incorporate both observational and
theoretical advances into our simple model, we do not apply the usual
observational benchmark of a fixed formation epoch, but instead allow
for a range of formation redshifts corresponding to the theoretically
favored model of hierarchical structure formation.  In this case we
allow the formation redshift to be a function of the virial
temperature, and choose $z_F(T_{\rm vir})$ such that the probability
of forming a galaxy at $z_F$ is the same for all values of $T_{\rm
vir}$.  Throughout this paper, the formation redshift corresponds to
the epoch of maximum gas infall.

We restrict our attention to the currently favored $\Lambda$CDM
model of structure formation, which is described in detail in
Eisenstein \& Hu (1999).  In this model, objects that are equally likely
to form have the same value of $\sigma (R)\,D(z_F)$, where $\sigma (R)$
is the amplitude of mass fluctuations inside a sphere of radius $R$
extrapolated linearly to $z=0$, given by their eq.\ (34) and (35), and
$D$ is a ``growth factor'' which tracks the linear growth of
perturbations as a function of redshift, given by their eq.\ (10).
Based mainly on the most recent measurements of the Cosmic 


\centerline{\null}
\vskip+3.3truein
\includegraphics{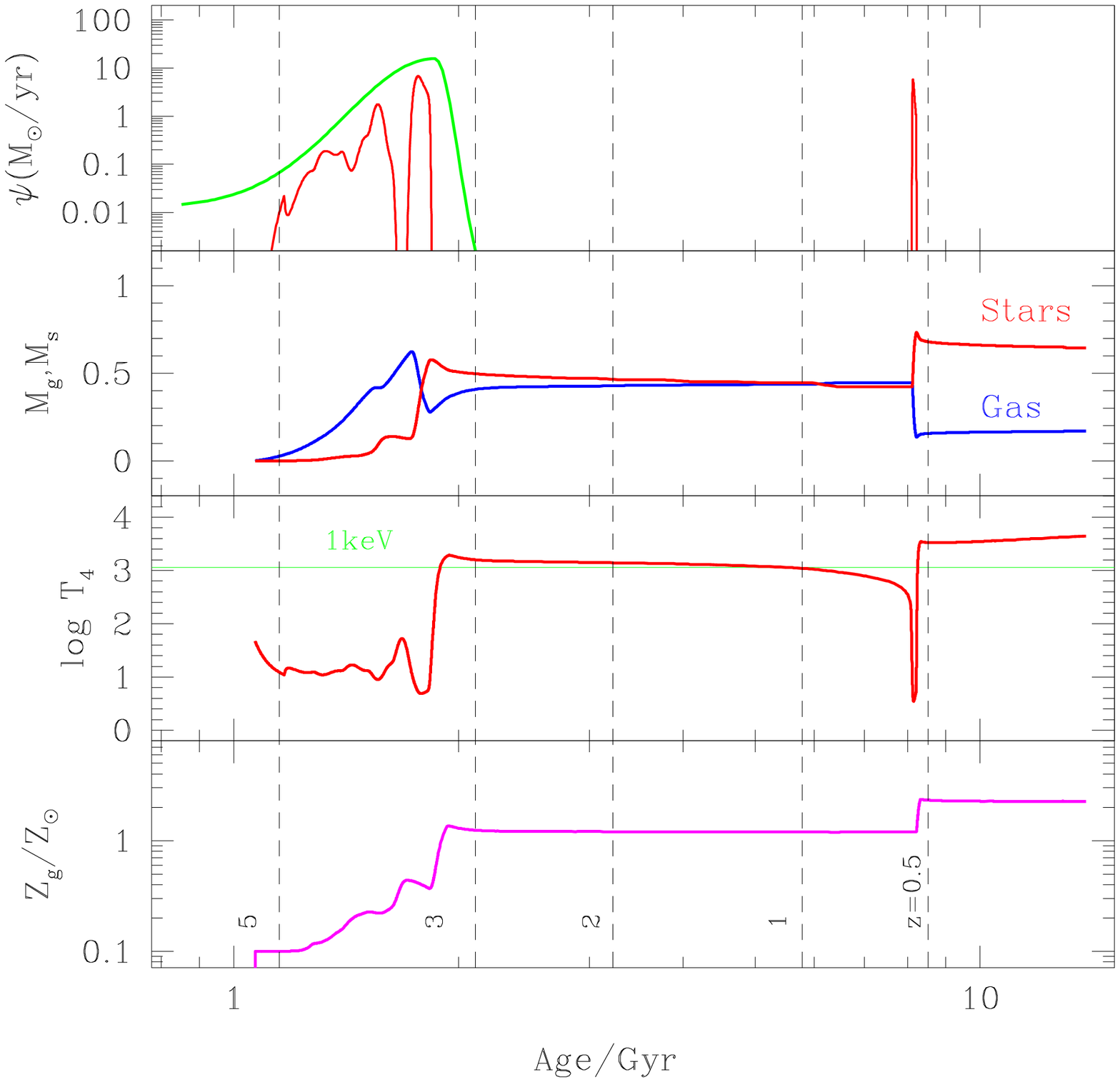}
\figcaption[th2.eps]{
Star formation history of a faint elliptical galaxy, 
which corresponds to $T_0=45$; $n_0=0.8$; $\tau_f=0.2$~Gyr; 
$z_F=3.9$; $\bout=0.5$. The photometry at zero redshift 
$U-V=1.2$; $V-K=2.9$ corresponds to a faint $M_V=-18$ elliptical 
(i.e. $\log\sigma\sim 1.8$) in a local cluster.
\label{fig:small_ellip}}
\vskip+0.2truein


\noindent Microwave
Background (Balbi et al.\ 2000; Netterfield et al.\ 2001; 
Pryke et al.\ 2002), we consider a cosmological model in which
$\Omega_m = 0.35$, $\Omega_\Lambda = 0.65$, $\Omega_b = 0.06$,
$\sigma_8 = 0.87$, $\Gamma = 0.18$, $n = 1$, and $h = 0.65$, where
$\Omega_m$, $\Omega_\Lambda$, and $\Omega_b$ are the total matter,
vacuum, and baryonic densities in units of the critical density,
$\sigma_8$ is $\sigma(r)$ at the 8$h^{-1}$ Mpc scale, $h$ is the
Hubble constant in units of 100 km s$^{-1}$ Mpc$^{-1}$, and $n$ is the
``tilt'' of the primordial power spectrum, where $n = 1$ corresponds
to a scale-invariant spectrum.  The age of the universe in this
cosmology is $\sim$ 14 Gyr.

Using this model, we select a range of objects and determine $z_F$ and
$T_4$ parametrically as a function of total mass, which is not itself
observable.  To calculate $z_F$, we compute the radius of the
perturbation, $R$, as $0.95 \Omega_m^{-1/3} M_{12}^{1/3}$
$h^{-1}$~Mpc, where $M_{12}$ is the mass in units of $10^{12} h^{-1}
M_\odot,$ and then invert $\sigma(R)D(z_F)={\rm const}$.  Similarly,
we calculate $T_4$ from a standard model of collapse and
virialization, which has been shown to be in good agreement with more
detailed numerical simulations (See e.g., Eke, Cole \& Frenk 1996). In
this case
\be
T_4 =  90 {\rm K} \,
M_{12}^{2/3} \, (1+z_F) \,
[{\Omega_m}+ \Omega_\Lambda (1+z_F)^{-3}]^{1/3}.
\label{eq:tvirialize}
\ee
Finally, we select $\sigma(R)D(z_F)$ such that the largest objects we
consider --- for which $T_{4} = 300$ --- 
form at a redshift of $z_F=2$, leading to a range of formation redshifts
from 2 to 4.4 (at $T_4 = 30$).  Note
that in this case ellipticals are formed from the so-called ``rare
peaks'' at redshifts which the mean amplitude of mass fluctuations,
$\sigma(R)D(z_F),$ is $2.5$ times smaller than the critical value for
collapse and virialization.  This is consistent with hierarchical
models which predict ellipticals to form at the peaks of the density
distribution due to their clustering properties and the fact that such
perturbations are likely to undergo a ``major merger'' (see eg.,
Kauffmann \& Charlot 1998).



 \begin{table*}
   \begin{center}
   \label{table-1}
   \vskip+0.1truein
   \caption{Mass-independent model parameters}
   \begin{tabular}{l|cr}\hline\hline
   Parameter & Symbol & Value\\
  \hline
   Schmidt-law index     & ${\cal N}$         & $1.5$ \\
   Pre-enrichment        & $Z_0$       & $Z_\odot /10$\\
   SN Temperature        & $T_{{\rm sm},4}$  & $3\times 10^5$\\
   SN Delay              & $\tau_{\rm sn} $ & $100$~Myr\\
   SF Efficiency         & $\ceff^0$   & $100$ \\
   Efficiency Steepness  & $\alpha$    & $3$\\
  \hline\hline
   \end{tabular}
   \end{center}
 \end{table*}

\bigskip

\section{Intermittent Star-Formation in Elliptical Galaxies}

\subsection{Global Parameters and Features of Feedback Models}

In this section, we describe the main features of our model and the
physics on which they depend.  In Table~1 we list the model parameters
that are independent  of mass.  The first two of these ${\cal N}=1.5$
and $Z_0 = Z_\odot /10$ are set to canonical values consistent with a
variety of observations.  The supernovae parameters $T_{{\rm sn},4}$
and $\tau_{\rm sn}$, on the other hand, are based on simple estimates
of the thermal energy released by each SNe and the time delay for this
energy to be transferred to the interstellar medium of the galaxy.
Finally the thermal SF efficiency $\ceff^0$ and steepness $\alpha$ are
arbitrary parameters in our model, whose values we have fixed
after an extensive search of parameter space.

Having established these values, we show in Figure~\ref{fig:big_ellip}
the model prediction for the evolution of a massive elliptical galaxy
($n_0=2$; $\bout =0$; $T_0=300$; $z_F=2$).  This object undergoes a
number of bursts of star formation at roughly constant time intervals
from its formation redshift of $z_F=2$ until the temperature of the
ISM rises above $\sim 1$~keV.  Notice that the feedback cycles
generated by our model are quite punctuated, with the gas remaining at
high temperatures for the majority of their evolution, interrupted by
brief intervals of ``catastrophic'' cooling that result in new bursts
of star formation.  Each burst is followed by a new generation of
supernovae, which quickly heats the ISM back up to the high
temperature state, where it remains quiescent until the next epoch of
cooling.

Each phase of this feedback cycle is dependent on the physical
parameters of our models.  In most cases, this dependence is quite
straightforward and can be understood by simple scaling arguments,
despite the fact that a more complicated numerical approach is
required to study the evolution in detail.  Before turning our
attention to the observable consequences of our modeling, we first
review these dependencies, comparing the properties of the large
galaxy shown in Figure \ref{fig:big_ellip} with an example of a much
smaller object with parameters ($n_0=0.8$; $\bout =0.5$; $T_0=45$;
$z_F=3.9$) in Figure \ref{fig:small_ellip}.

Both of these galaxies start at their virial temperatures, and remain
there for a significant period of time, before undergoing a first
burst of star-formation. At high temperatures $\gtrsim 10^5 K$, the
cooling function $\Lambda$ roughly scales as $1/T$, thus the cooling
time of the gas $\tau_{\rm cool}=kT/n\Lambda$ is approximately
$\propto T^3,$ as $n \propto 1/T$ during the cooling phase.  This
strong dependence on temperature means that the main barrier to
star-formation in these objects is the cooling time {\em at the
highest temperature in the feedback cycle}.  Once the galaxy cools to
temperatures $T\sim 10^4$K, the sharp decline of the cooling function
halts the thermal run-away, while at the same time star-formation
becomes much more efficient.  Note that the strong scaling of the
cooling function with temperature means that this lower value is
nearly constant over the full range of galaxy masses.  It is during
this stage that almost all stars are formed, resulting in a starburst
with total star formation rates of order 1000~$M_\odot$/yr in massive
galaxies and 10~$M_\odot$/yr in low-mass systems.  The galaxy remains
in a low-temperature state until the first generation of supernovae
explode and transfer their thermal energy to the gas.  Thus each
starburst results in a total fraction of stars created, which is
roughly $\ceff \times (\tau_{\rm sn} + \tau_{m=100 M_\odot}) \times
\mu_g(t_{\rm SB})$, where $\mu_g(t_{\rm SB})$ is gas mass at the start
of the starburst.  Similarly, the post starburst temperature is simply
proportional to the number of stars formed times the energy
transferred to the gas: $E \propto (1-\bout ) T_{\rm sn} \times \ceff
\times (\tau_{\rm sn} + \tau_{m=100 M_\odot}) \times \mu_g(t_{\rm
SB})$.

From these scalings, it is easy to understand why smaller galaxies
oscillate more frequently than those with larger masses.  As the
energy per supernova, star formation efficiency, and $\tau_{\rm sn}$
are fixed for all models, the supernova energy generated in each model
varies only by a factor $(1-\bout ) \times \mu_g(t_{\rm SB})$.  But
for each starburst, the ratio of the initial and final gas mass is
fixed and the post starburst temperature scales simply as $\propto
(1-\bout )$.  Thus the period between starbursts can be estimated as
approximately $\propto T^3\propto (1-\bout )^3$.

Comparing Figures~$1$ and $2$, which give the star formation histories
of two early-type galaxies with very different masses, we see that the
ratio of gas outflow fractions is $2$, implying that the final
bursts of star formation will take place approximately $8$ times later
in the high-mass galaxy.  Thus the last early burst in the large
galaxy takes place after a delay of approximately 1 Gyr, while the
last early burst in the $T_4=45$ model is partially blended with the
peak before it, with a relative separation of approximately $1/8$~Gyr.
Similarly, after the early burst stage, the smaller galaxy reaches a
temperature of $1.5 \times 10^7$K, and cools to form a secondary late
burst of star formation after approximately $7$ Gyrs.  On the other
hand, the large galaxy reaches a temperature of $\approx 3 \times
10^7$K after the initial burst phase, corresponding to a cooling time
of $\sim 50$ Gyrs, and therefore no late bursts appear in this model.

It is important to note that the relationship between mass and
metallicity in our models arises not only from a differential $B_{\rm
out}$ as in Ferreras and Silk (2000b), but from an interplay between
the multiple bursts and the infall timescale.  This aspect of our
model is independent of the late time star formation properties of
these galaxies.  Thus it is conceivable that the early time
star-formation history of the galaxy may have been somewhat different
(perhaps due to the merger of two disks for example) while still
preserving the star formation cycles and scalings seen in our models
at late times.  The metallicity-mass relation is dis-


\centerline{\null}
\vskip+3.3truein
\includegraphics{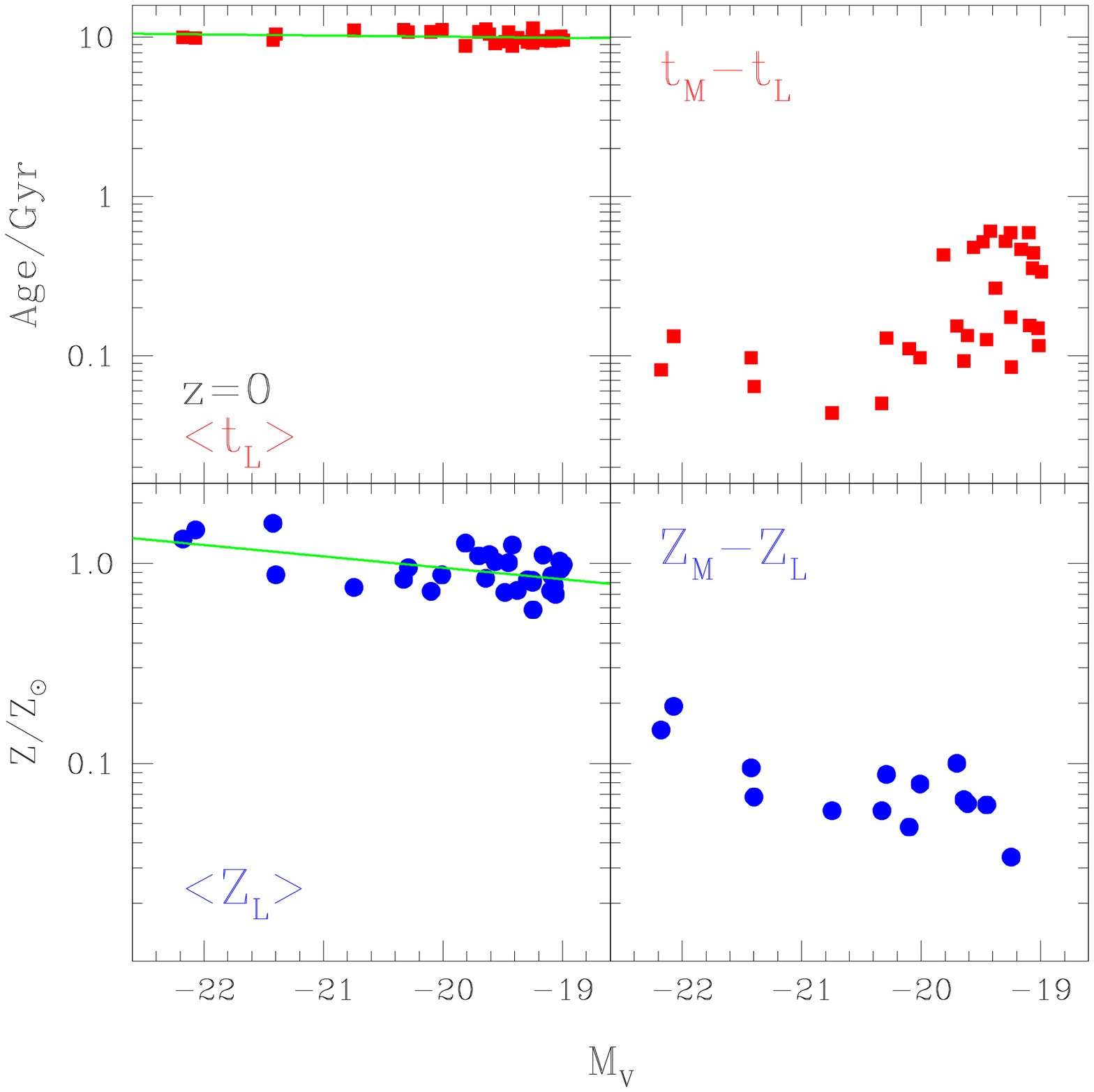}
\figcaption[th3.eps]{
Average $V$-band luminosity-weighted age and metallicity 
({\sl left}) and its difference with a mass-weighted average 
({\sl right}) at zero redshift. The lines in the left panels give
a least squares fit to the age-mass and metallicity-mass relations. 
The slopes of these linear fits are $0.02$ and $0.12$, respectively. 
\label{fig:age1}}
\vskip+0.2truein


\noindent cussed in more detail in \S4.

Finally, in Figures 1 and 2, the thick lines represent an estimate as
to the physical star formation rate in these objects.  This is likely
to be a smoothed version of the theoretical curve, as our model
assumes a point-like galaxy, whereas a more realistic model including
spatial information would have bursts of star formation happening at
slightly different epochs in different regions of the galaxy.  This
would result in a more continuous global star formation rate during
the first stage, while preserving the punctuated bursts at 
later times.

This raises the question as to whether the interactions between areas
of the galaxy in which star formation is out of phase may be able to
suppress the cycles seen in our one-zone models.  Here the important
issue is whether SNe are able to heat a region of space and suppress
star formation fast enough that spatial variations between regions
become important.  In our one-zone models, we assume a typical delay
of 100 Myrs for SN to percolate through the galaxy, dumping a fraction
$B_{\rm out}$ of their energy into the ISM.  Thus we expect a one-zone
description to be applicable if the galaxy is relatively homogeneous on
scales comparable to 100 Myrs times the sound speed of the gas.  A
simple estimate of this speed gives $c_s \approx 10 T_4^{1/2}$ km/s,
our one-zone approach is equivalent to assuming that the galaxy is
relatively smooth on scales above $\tau_{\rm SN} c_s \approx 1
T_4^{1/2}$ kpc, a reasonably large scale even at $T_4 = 1.$ Note
however that our value for the SNe delay time is somewhat arbitrary and
chosen empirically to reproduce the observed properties of
elliptical galaxies.  Thus while our model is a plausible description
of a more complete treatment, a dedicated numerical study would be
necessary to more firmly establish the relevant thermalization time
in our models, and whether $c_s \tau$ is large enough that spatial
variations can be ignored.

\subsection{Mass Dependent Parameters and Average Properties}


\centerline{\null}
\vskip+3.3truein
\includegraphics{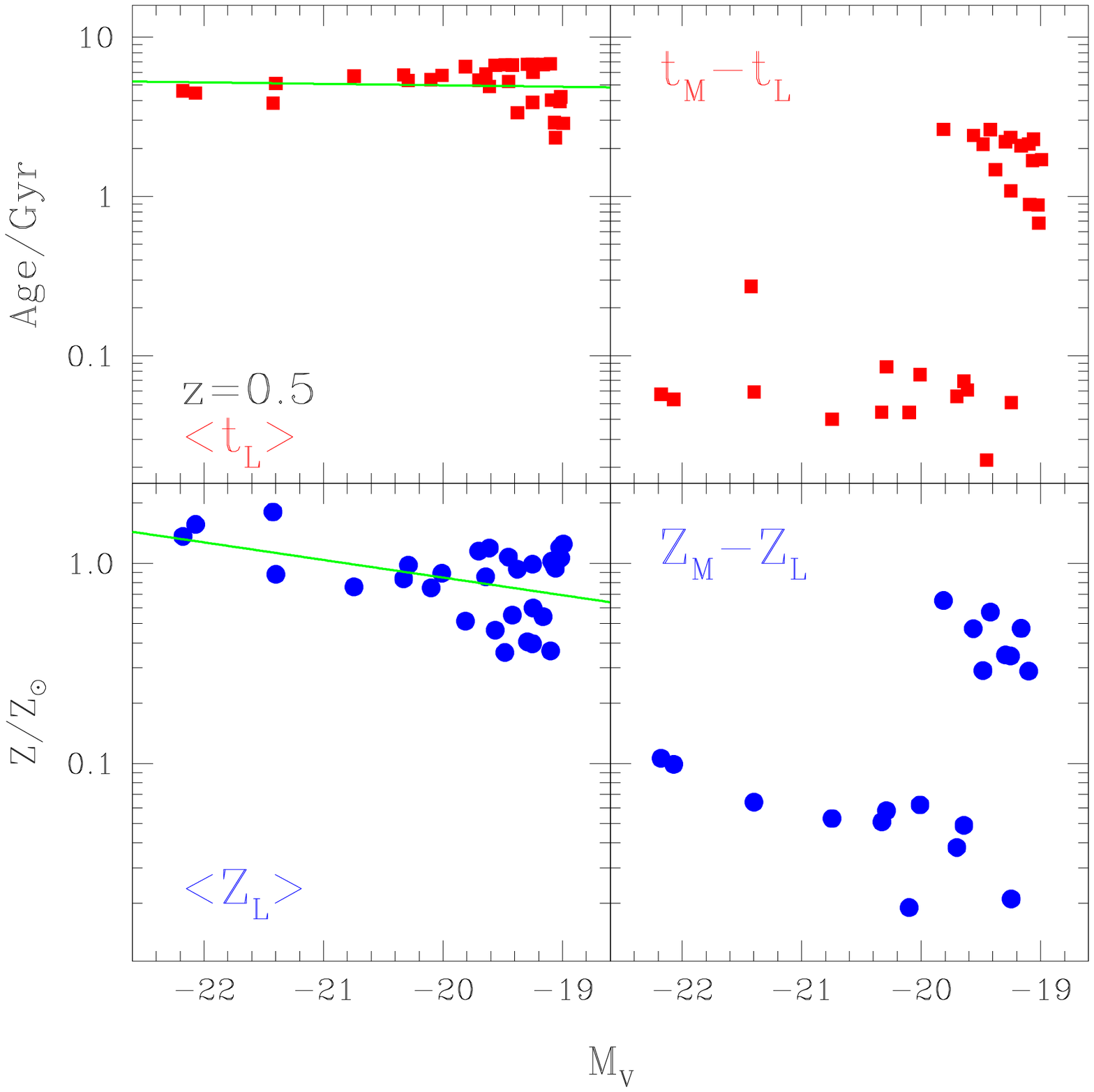}
\figcaption[th4.eps]{
Same as figure~\ref{fig:age1} for a moderate redshift 
($z=0.5$) cluster. The slopes of 
the linear fits to the age-mass and metallicity-mass relations 
are $0.02$ and $0.20$, respectively.
\label{fig:age2}}
\vskip+0.2truein

Having examined the main features of our models, we now turn our
attention to constructing a representative sample of elliptical
galaxies over a range of masses.  In Table~\ref{table-2} we list the
parameters of our model that depend on age and indicate their
scaling with mass and formation redshift.

The virial temperature $T_{\rm vir}$ simply scales as $G M^{2/3}
\rho^{1/3}$, where $\rho$ is the mean cosmological gas density at the
time of formation.  Similarly the initial number density of the gas at $10^4
K$, $n_0 \propto \rho T_{\rm vir}$ as $n \propto 1/T$ during the
cooling phase.  The infall timescale, on the other hand, scales as the
radius of the halo $\propto M^{1/3} \rho^{1/3}$ divided by the virial
velocity $\propto T_{\rm vir}^{1/2}.$ Finally, we assume a simple
power law dependence with mass for the fraction of gas ejected in
outflows, $\bout$.  The most massive galaxies are expected to have
negligible outflows ($\bout\sim 0$), while the low mass galaxies
should have higher outflow fractions that will reduce the average
metallicity of the stellar populations, thereby blueing their
colors. We use the color of faint ellipticals in local clusters to
constrain $\bout$ in these systems, predicting roughly $\bout\sim 0.6$
for $M_{10}\sim 5$.  Thus the power law index over the mass range
$5<M_{10}<300$ is $\sim -0.3$.

Using the mass-independent parameters described in Table~\ref{table-1}
as well as the scaling of the mass-dependent parameters shown in
Table~\ref{table-2}, we can generate a sample of galaxies starting
from a luminosity function.  We estimate the relative number densities
of these objects using the NIR $H$-band luminosity function of the
Coma cluster (De~Propris et al.\ 1998), which provides a good estimate
of the underlying mass function, and arrive at a final sample of
galaxies studied in detail below.

Figures~\ref{fig:age1} and \ref{fig:age2} show mass and $V$-band
luminosity-weighted averaged ages and metallicities of the objects, 
defined as:
\be
<t_M> = \int_0^{t(z)}dt\, t\psi(t) {\Big /} 
        \int_0^{t(z)}dt\, \psi(t),
\ee
\be
<Z_M> = \int_0^{t(z)}dt\, Z\psi(t) {\Big /} 
        \int_0^{t(z)}dt\, \psi(t),
\ee


\centerline{\null}
\vskip+3.3truein
\includegraphics{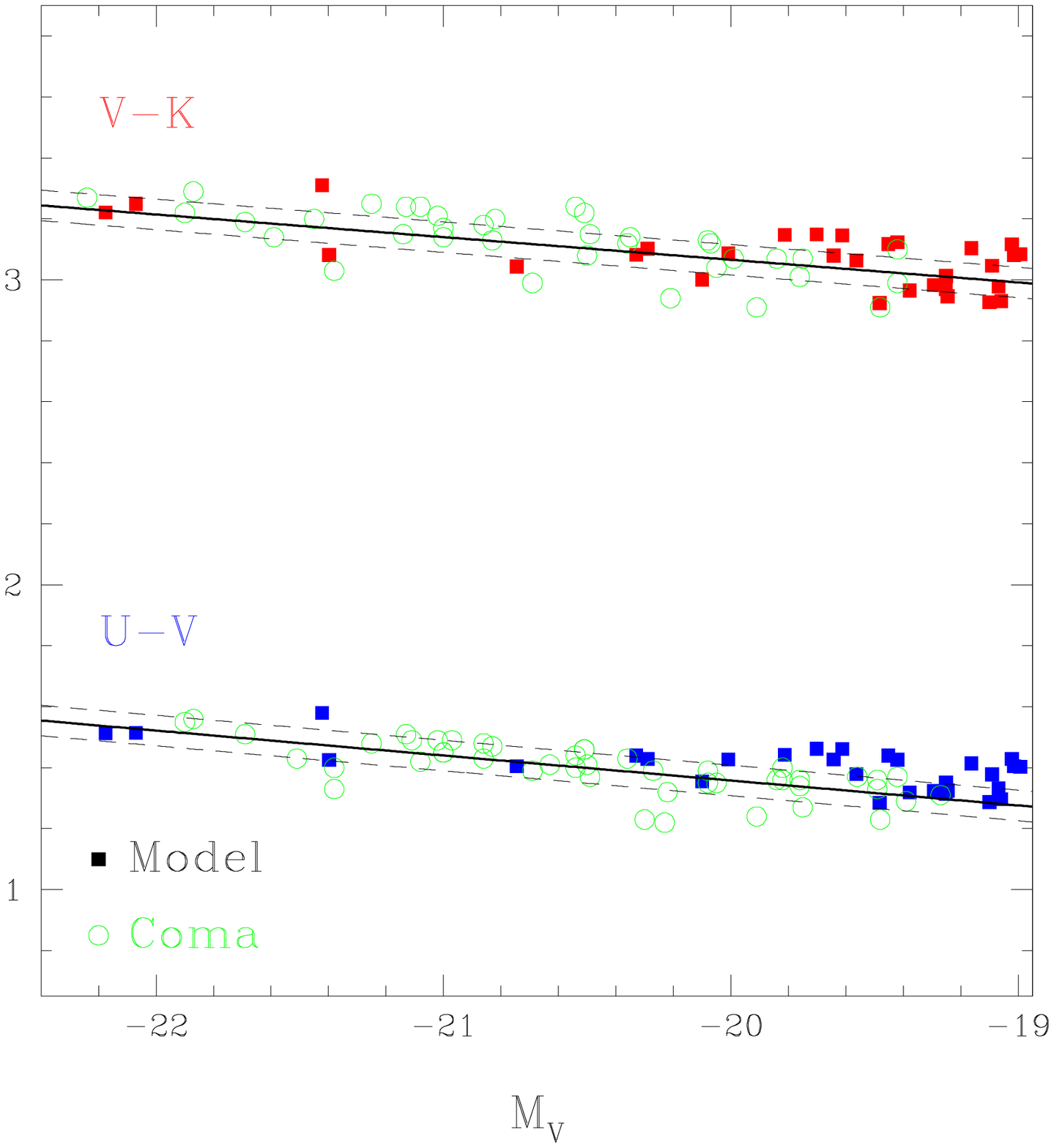}
\figcaption[th5.eps]{
Color-magnitude relation in optical and near-infrared passbands 
for a local cluster. The filled squares are our model predictions. 
The hollow circles are Coma cluster ellipticals observed by  
Bower et al.\ (1992). The best linear fit to the observations is
given by the solid line, and the $\pm 0.05$ mag scatter found in these
galaxies is plotted as a dashed line.
\label{fig:cmz0}}
\vskip+0.2truein


\be
<t_L> = \int_0^{t(z)}dt\, t\Upsilon(t)^{-1}\psi(t) {\Big /}
        \int_0^{t(z)}dt\, \Upsilon(t)^{-1}\psi(t),
\ee
\be
<Z_L> = \int_0^{t(z)}dt\, Z\Upsilon(t)^{-1}\psi(t) {\Big /}
        \int_0^{t(z)}dt\, \Upsilon(t)^{-1}\psi(t),
\ee
where $\Upsilon (t)$ is the $V$-band mass-to-light ratio as a function
of stellar age, and $t(z)$ is the age of the universe at the
redshift at which it is observered. Comparing these results, it
is clear that the ages of our model galaxies at moderate redshift are
strongly dependent on which type of average is applied.  While the
mass-weighted values provide the most meaningful summary of the
overall star-formation history of the object, both photometric and
spectral observations only give estimates of the luminosity-weighted
age and metallicity. Thus, late episodes of star formation skew the
integrated spectral energy distribution to younger ages, because of
the very small mass-to-light ratios of high-mass main sequence stars.

Figure~\ref{fig:age1} shows that at zero redshift, early-type galaxies
can be described by a metallicity sequence roughly between
1.5$Z_\odot$ for the most massive galaxies, down to $0.5 Z_\odot$ for
the faintest ones, in agreement with observations of ellipticals in
local clusters (Kuntschner \& Davies 1998; Kobayashi \& Arimoto 1999).
Our model gives a power law fit between mass and metallicity of
$Z\propto M^{0.15}$ at zero redshift.  The slope of the age-mass
relation is nearly flat both at zero and moderate ($z=0.5$) redshift
($t\propto M^{0.02}$), although the latter presents a significantly
larger scatter. While there is little difference between mass and
luminosity-weighted ages at zero redshift, Figure~\ref{fig:age2} shows
significant differences at $z=0.5$.  The mass-metallicity correlation
in this figure is clearly defined and has a similar power law index:
$Z\propto M^{0.20}$. However, there is an increased scatter in the
luminosity-weighted ages and metallicities, especially at 


\centerline{\null}
\vskip+3.3truein
\includegraphics{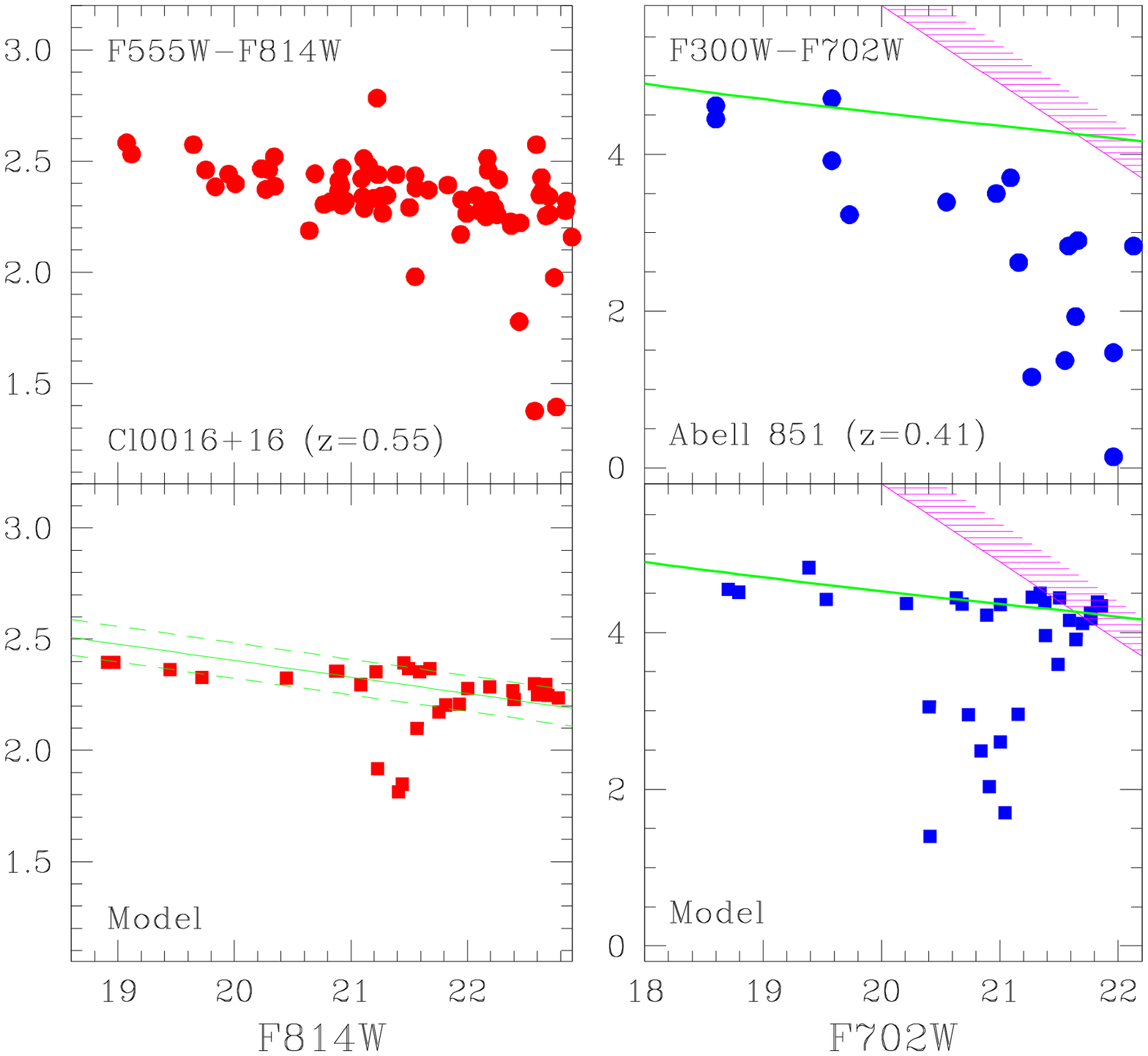}
\figcaption[th6.eps]{
Rest frame $U-V$ ({\sl left}) and NUV$-$ optical ({\sl right})
color-magnitude relations predicted at redshifts of $z=0.55$ and $z=0.41$. 
The filled squares are the model predictions,
whereas the hollow circles show the data from cluster
Abell~851 ($z=0.41$) explored in the NUV by Ferreras \& Silk (2000a)
and Cl0016+16 ($z=0.55$) observed by the MORPHS collaboration 
(Ellis et al.\ 1997). The shaded areas in the panels on the right give 
the $4\sigma$ detection limit from the shallower F300W images. 
The solid line in the panels on the right gives the prediction of a single-age 
metallicity sequence formed at $z_F=10$ using the color-magnitude 
of the Coma cluster as constraint. The solid
and dashed lines in the bottom left panel is the fit and scatter,
respectively, to the color magnitude relation in the elliptical
galaxies of cluster Cl0016+16.
\label{fig:cmz}}
\vskip+0.2truein


\noindent the faint
end. This offset can be as large as a few Gyr in age, especially for
galaxies with secondary bursts occurring within the previous
1~Gyr. This occurs since even small amounts of young stars can boost
the luminosity mainly in the NUV, thereby reducing the
luminosity-weighted age.


 \begin{table*}
   \begin{center}
   \caption{Mass-dependent model parameters}
   \label{table-2}
   \vskip+0.1truein
   \begin{tabular}{l|ll}\hline\hline
   Parameter & Symbol & Scaling\\
  \hline
   Virial temperature & $T_{{\rm vir},4}$ & $\propto M_{10}^{2/3}(1+z_F)$\\
   Initial number density at $T_4=1$ & $n_0$ & $\propto T_{{\rm vir},4}(1+z_F)^3$ \\
   Infall timescale   & $\tau_f$ & $\propto (1+z_F)^{-3/2}$\\
   Outflow gas fraction   & $\bout$ & $\propto -0.3\log M_{10}$\\
  \hline\hline
   \end{tabular}
   \end{center}
 \end{table*}

\section{Optical and Near-Ultraviolet Color-Magnitude Relations
	in Ellipticals}

For a given choice of parameters, our model predicts a chemical
enrichment track that determines the ages and metallicities of the
stellar populations. These star formation histories can be convolved
with models of simple stellar populations to get the
spectrophotometric properties of the model galaxies. To do this, we
use the latest population synthesis models of Bruzual \& Charlot (in
preparation) and assume a fixed baryon to total mass ratio of
$\Omega_b/\Omega_m \sim 1/6$ in order to obtain absolute luminosities.

Figure~\ref{fig:cmz0} gives the optical and near-infrared (NIR)
color-magnitude relation predicted for a local cluster. The model
galaxies are shown as filled squares and are compared with the
precision photometry of Coma ellipticals observed by Bower, Lucey \&
Ellis (1992), plotted as hollow circles.  The solid and dashed lines are the
best fits and scatter to the observations, respectively. The
remarkable agreement shows that our $\bout$ scaling reproduces well
the expected mass-metallicity or mass-luminosity relationship (as
shown in figure~\ref{fig:age1}) that explains the color range in
early-type galaxies. Notice that the $V-K$ color departs towards the blue
at the faint end more than the $U-V$ color. This has to do with the
higher age sensitivity of optical colors with respect to near-infrared
passbands. The $U$ and $V$-bands will be more affected by recent
episodes of star formation than $K$-band, so that $V-K$ appears bluer
than $U-V$ at the faint end at which late bursts of star formation
occur.

Figure~\ref{fig:cmz} shows the rest-frame 2000\AA\ near-ultraviolet
(NUV) and optical color-magnitude relation predicted for two clusters
at moderate redshift: Abell~851 ($z=0.4$; {\sl right}) whose optical
and NUV color-magnitude relation was recently analyzed by Ferreras \&
Silk (2000a), and Cl0016+16 ($z=0.55$; {\sl left}) whose detailed
optical color-magnitude relation can be found in Ellis et al.\ (1997). 
The shaded area gives the $4\sigma$ detection limit for
the shallower F300W images. The solid and dashed lines show the fit
and scatter to the observed data in cluster Cl0016+16. 

These two color-magnitude relations paint two very different pictures
of a similar cluster at moderate redshift.  The optical colors 
shown in the panels on the left do not present a large
scatter.  This is consistent with the observations of $z \lesssim 1$
clusters by Stanford, Eisenhardt, \& Dickinson (1998), which they
interpreted as an absence of star formation at redshifts below $2-3$.

The near-ultraviolet color-magnitude relation showed in the 
panels on the right, however, point to a more involved SF history.  
In this case
the blue band maps a rest frame spectral window around 2000\AA\ in
which young main sequence A-type stars contribute significantly,
making these colors much more sensitive to recent star formation.  Thus
the observed scatter with respect to the prediction for old stellar
populations formed at redshifts $z\sim 5-10$ (thick line) is a
telltale sign of recent star formation. This is particularly clear
as the lookback time of this cluster allows us to safely ignore the
contribution to the F300W flux from evolved, core helium-burning 
stars, whose contribution would peak around
1500\AA. Hence the combination of NUV and optical colors of
early-type galaxies at moderate redshifts is a valuable
observable to infer the recent star formation histories of these
systems. Yet detailed analyses of NUV$-$optical color-magnitude
relations at moderate redshift are scarce (Buson et al.\ 2000; Ferreras
\& Silk 2000a) and more work is needed in this direction.

Mass-to-light ratios are also useful stellar clocks although their
observed estimates are highly model dependent. Figure~\ref{fig:ML}
shows the model predictions against total mass in units of
$10^{10}M_\odot$. Galaxies with late bursts depart from the ``red
envelope'' since their young stars contribute significantly to the
total $V$-band luminosity, even if they make up only a small fraction
of the total stellar mass.  Here the solid line is the corrected fit
to the observed data from Mobasher et al.\ (1999; stars) and Pahre
(1999; hollow squares) which gives $M/L_V\propto M^{0.24}$.  The
correction involves a slightly nonlinear power law dependence between
the stellar mass $M_s$ and the total mass $M\propto M_s^{1.2}$ as
described in Ferreras \& Silk (2000b).

\section{X-ray Emission from Early-Type Galaxies}

The model presented in this paper gives the thermal state of the ISM a
central role in the star formation history of elliptical
galaxies. Even though our model is a rather simple 1-zone system, we
can use this information to predict the X-ray luminosity from the gas
and its correlation with the optical luminosity, which traces the
stellar component.

The X-ray luminosity of the galaxy can be approximated as
arising purely from Bremsstrahlung emission from free-free electron 
interactions.  In this case 
\be
L_X \approx  2 \times 10^{-25} n_e N_e T_4^{1/2} 
\left(e^{-\frac{E_{\rm max}}{k T}}- e^{-\frac{E_{\rm min}}{k T}} \right) 
{\rm ergs} \, {\rm cm} \, {\rm s}^{-1},
\ee
where $E_{\rm max}$ and $E_{\rm min}$ are the maximum and minimum
energies of the observed X-ray band.  Due to line emission, this
equation is only valid at temperatures much greater than $T_4=1$, but
this is a good assumption as the X-ray emission from cold objects is
negligible. A larger source of error, in fact, is that we only include
X-ray emission from the gas and not from X-ray binaries, which may
contribute significantly.

Rewriting the number density in terms of 
temperature and $n$, using $N_e = 1.5 \times 10^{66} M_{10}$ 
in this cosmology, and restricting
our attention to the $0.5$ to $2.4$~kev energy band, we find 
\be
L_X \approx 3 \times 10^{41} \, M_{10} \, n \, {\rm cm}^3 \, \mu_g \, 
T_4^{1/2} \left(e^{-\frac{530}{T_4}}- e^{-\frac{3400}{T_4}} \right) 
{\rm ergs} \, {\rm s}^{-1}.
\ee
The filled squares in 
Figure~\ref{fig:Xray} show the predicted $L_X$ vs $L_B$ correlation
in early-type galaxies following this equation.
The hollow circles are taken from O'Sullivan et al.\
(2001) who compiled a sample of 401 early-type galaxies. The solid
line gives the best fit to the observed data, excluding bright cluster
and group galaxies as well as AGNs. Our model is in good agreement,
showing that the trend of the $L_X:L_B$ correlation in elliptical
galaxies can be mostly explained by X-ray emission from the hot
interstellar medium.  While the overall scatter in the data exceeds
that in the models somewhat, this is not surprising, as the
observational values are subject to additional variations not only due
to the X-ray binaries as described above, but perhaps also due to
environmental effects (Mathews \& Brighenti 1998).

We can also use a simple scaling argument
to estimate sizes from the galaxy mass ($M_{10}$) and number density
($n_0 \approx n T_4$). In this case
\be
R \sim 5~{\rm kpc} \left( \frac{M_{10} \mu_g}{n_0m_{0.7}}\right)^{1/3}.
\label{eq:size}
\ee
For our model galaxies the inferred sizes range from $8$ to $30$~kpc,
which is a reasonable estimate of the core X-ray sizes of ellipticals
(Mathews \& Brighenti 1998). High-resolution deep X-ray 
images of ellipticals over a large mass range would be 


\centerline{\null}
\vskip+3.3truein
\includegraphics{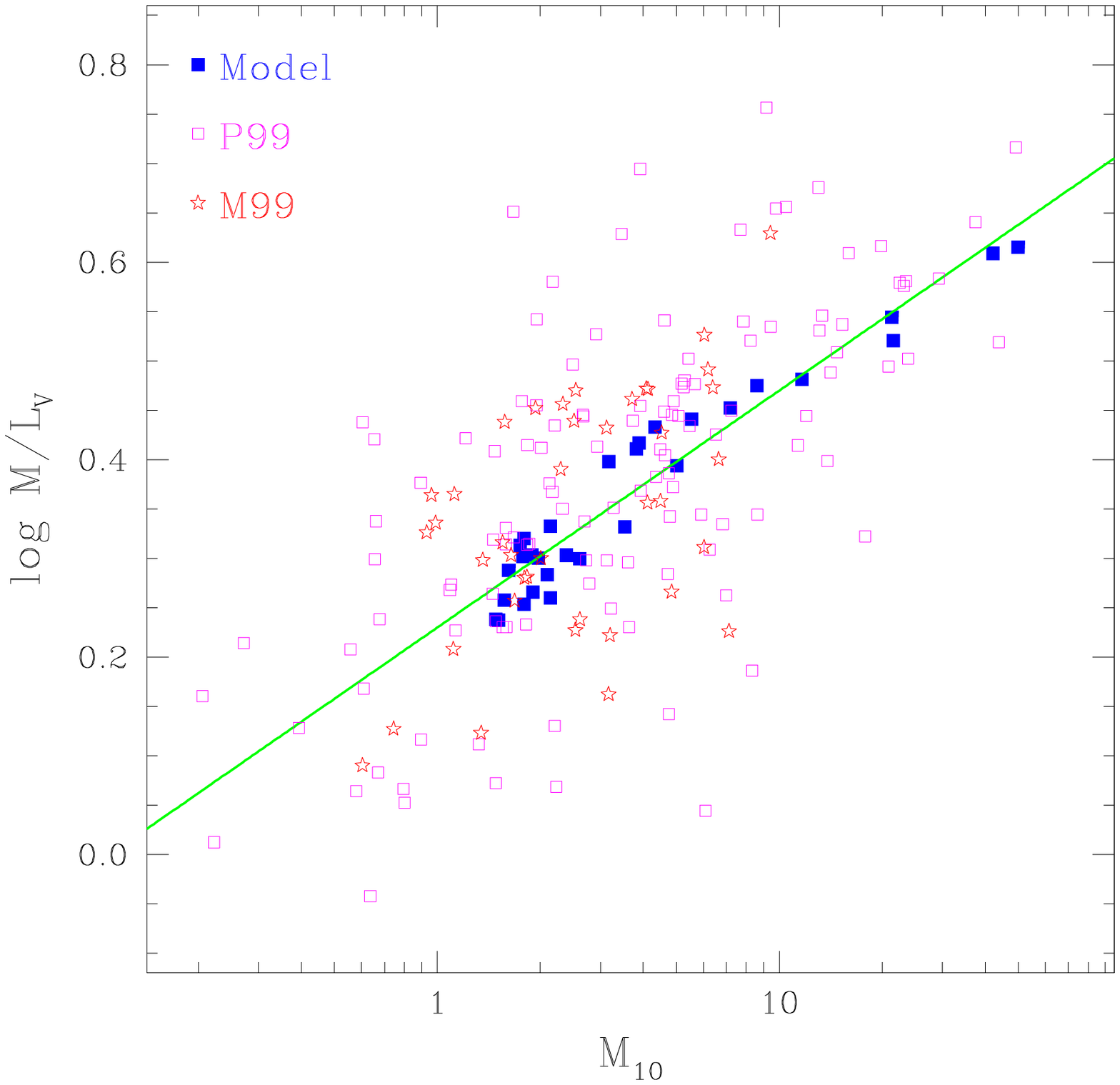}
\figcaption[th7.eps]{
Correlation between the {\sl stellar} 
$V$-band mass-to-light ratio as a function of total mass in 
units of $10^{10}M_\odot$. The model data (filled squares) 
have been corrected for the offset between the total and the 
stellar mass content in elliptical galaxies (Ferreras \& Silk 2000b). 
The observations are from Pahre (1999; P99) and Mobasher 
et al.\ (1999; M99).
\label{fig:ML}}
\vskip+0.2truein


\noindent 
needed to determine the scaling of galaxy X-ray size with 
optical luminosity, which could quantify the contribution to the 
X-ray brightness from the hot interstellar gas in elliptical galaxies.


\section{Toward a More Complete Picture of Feedback in Early-Type Galaxies}

While our simple model of thermal feedback yields a promising
explanation for many of the observed properties of early-type
galaxies, it naturally raises two questions.  As secondary starbursts
may also arise from more complicated gas accretion histories that the
one consider here, one is left with the question of the what types of
observations will be most important in distinguishing between our
scenario and other possibilities.  Secondly, from a theoretical point
of view, it is clear that intense bursts of supernovae will 
deposit both energy and momentum over a much larger area than the ISM
of the galaxy itself.  This raises the question as to the uniqueness
of our single zone thermal model, and whether similar feedback loops
can also arise in the interplay between the galaxy and its
environment.  In this section we address each of these questions in
turn.

\subsection{Observational Implications}

The elliptical SF histories described in this paper are much more
complicated than those of more usual models which assume a short epoch
of star formation resulting in stellar populations with well-defined
ages and metallicities.  Although our predictions for massive galaxies
are quite similar to such models, our lower mass systems experience
secondary episodes of star formation that affect their luminosities,
particularly at moderate redshifts. This is true independent of the
possibility of other sources of late star formation from
environmental effects.

Our conclusion clearly calls for more detailed analyses of elliptical
galaxies at moderate and high redshift.  It also gives an
additional mechanism 
for ``hiding'' high-redshift ellipticals 


\centerline{\null}
\vskip+3.3truein
\includegraphics{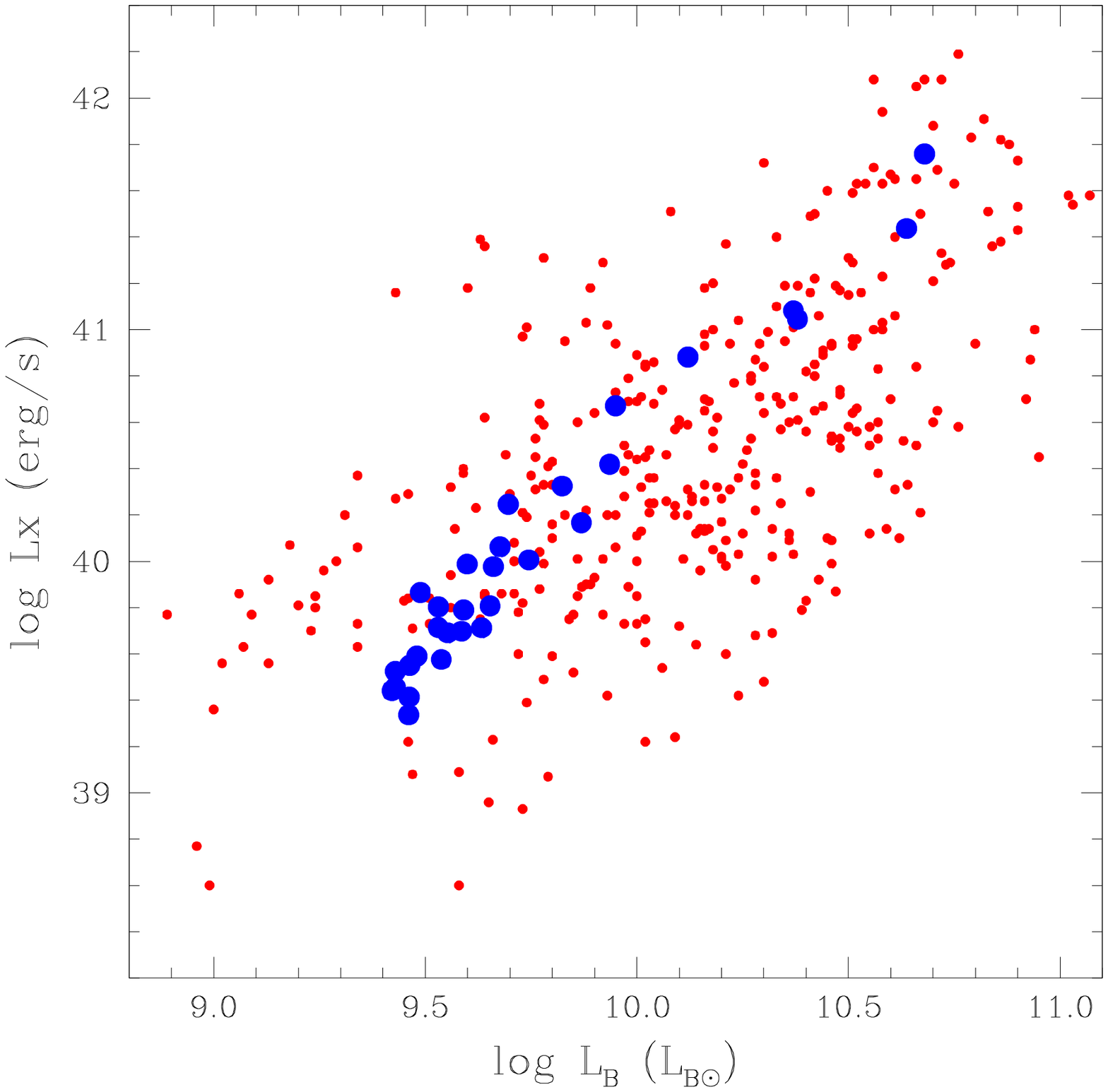}
\figcaption[th8.eps]{
Predicted correlation between the X-ray and optical luminosities
using a simple argument involving Bremsstrahlung emission from the electrons
in the hot gas component. The filled squares give our model predictions, 
whereas the hollow circles are from O'Sullivan et al.\ (2001). 
\label{fig:Xray}}
\vskip+0.2truein


\noindent from view.
In fact, this possibility is reminiscent of the the blue-nucleated
field early-type galaxies observed by Menanteau, Abraham \& Ellis
(2001) in the Hubble Deep Field, which may be examples of the late
bursts predicted by our model.  Similarly, the post-starburst E+A
galaxies found in clusters are suggestive of our results.

It is also interesting to note that the observed correlation between
abundance ratio enhancements and galaxy mass or luminosity (Kuntschner
\& Davies 1998; Trager et al.\ 2000) is indicative of extended star
formation in low mass galaxies, which allows the byproducts from
type~Ia supernovae to be locked into subsequent generations of stars.
Galaxies with central velocity dispersions around 50~km~s$^{-1}$, for
example, tend to have [Mg/Fe] ratios close to Solar, while values
around [Mg/Fe]$\sim +0.3$ are observed in massive ($\sigma_0\sim
400$~km~s$^{-1}$) galaxies.  Again, this hints at late bursts of star
formation.

While the short duration of our secondary bursts makes their direct
detection a complicated endeavor, observations of NUV colors may
provide a tractable approach to studying feedback.  As young
main-sequence A-type stars have strong Balmer absorption lines in the
NUV spectral window around 2000\AA, observations that map this
rest-frame wavelength can used to trace recent star formation.  Thus
while low numbers of cluster galaxies with strong emission lines are
seen at moderate redshift (Dressler et al.\ 1999), up to 20\% of such
galaxies feature post-starburst NUV spectra (Poggianti et al.\ 1999).
The large scatter at the faint end of the NUV minus optical
color-magnitude relation of Abell~851 (Ferreras \& Silk 2000a) is a
further indication of the sensitivity of this approach.  As the
thermal feedback mechanism presented in this paper is independent of
the environment, comparisons of these colors in the field, groups, and
cluster populations should provide a direct method of separating the
contribution of thermal feedback to star formation in spheroids from
other processes.

\subsection{Alternative Models of Feedback}

The model of feedback in elliptical galaxies we have examined in this
paper focuses on ISM heating, modulating star formation by changing
the condition of the gas within the galaxy itself.  Yet the presence
of a large number of supernovae following a burst of star-formation is
likely to also deposit significant amounts of energy and momentum into
the surrounding intergalactic medium.  In our thermal model, infall
onto the galaxy was fixed simply by the dynamical timescale of a
virialized cloud as a function of mass and redshift.  While this is a
reasonable approximation at early times, the rate of infall from the
intergalactic medium after the onset of star-formation may be slowed
due to collisions with supernova ejecta.  This raises the question as
to whether multiple starbursts can also arise in models that focus on
the interplay between the galaxy and its environment.

To study the features of such models, we can replace our
simple two-component model with a three-component model in which we
track stars, the gas contained within the galaxy, and a second
gas reservoir that represents the surrounding medium, which is 
condensing onto the galaxy.  In order to isolate the features of such
an approach from the thermal effects discussed above, we do not
attempt to track the temperature of the gas, but rather consider a
generic set of equations describing a broad class of models in which
infall is modulated by feedback.  In this case we find
\ba
\dot{\mu}_g &=&  \mu_r/\tau - \epsilon_{\rm mom} E - \ceff \mu_g +
(1-f)(1-\bout)E, \nonumber \\
\dot{\mu}_r &=& -\mu_r/\tau + \epsilon_{\rm mom} E + 
	f(1-\bout)E, \nonumber \\
\dot{\mu}_s &=& \ceff \mu_g,
\label{eq:momback}
\ea
where $\epsilon_{\rm mom}$ is a parameter of order 
$\sqrt{10^{51} {\rm ergs}/M_\odot}/v_{\rm vir} \approx 10/v_{100}$
that parameterizes the suppression of infall by supernovae,
$f$ is the fraction of ejected material that is recycled
into the infalling reservoir, and the infall is now
parameterized as  an exponential with a scale time $\tau$
in order to simplify the analytical arguments below.

With these equations in hand we now approximate the behavior of this
system using a ``post'' instantaneous recycling approximation in which
we model $E(t)$ considering the timescale ($\tau_2$) corresponding to
the lifetime of stars that play a role in stellar feedback.  In this
approximation we write the ejected fraction as
\begin{equation}
E(t)\sim \ceff f_{\rm SN} \Big[ \mu_g(t) - \dot{\mu}_g(t)\tau_2\Big],
\end{equation}
where $ f_{\rm SN}$ is the gas fraction returned by supernovae.  Note
that a good estimate of $\tau_2$ is $\sim 0.02$~Gyr (roughly the
lifetime of a massive star that undergoes core collapse, i.e.  $M\sim
10 M_\odot$), but we are also free to take $\tau_2 =0$, recovering the
more commonly studied instantaneous recycling approximation.

Rewriting these equations in terms of the infall timescale and
defining $s\equiv t/\tau_1$; ${\cal C}=\ceff\tau_1$; and $x\equiv
\tau_2/\tau_1$, we find
\begin{equation}
( 1-{\cal C} L \Big) \frac{d^2\mu_g}{ds^2} +
( 1+{\cal C} M ) 
\frac{d\mu_g}{ds} + {\cal C} N \mu_g =0,
\label{eq:harm}
\end{equation}
where
$L \equiv  f_{\rm SN} \, \epsilon' \, x$,
$M \equiv 1 \, +  \, f_{\rm SN} \, \epsilon' + (1-\bout ) \, f_{\rm SN} \, x$, 
$N \equiv 1 \, - \, (1-\bout )  \, f_{\rm SN}$,
and $\epsilon' = \epsilon_{\rm mom} \, - \, (1-f) \,(1-\bout ).$
Note that for any reasonable choice of parameters $\epsilon' > 0$ as
$\epsilon_{\rm mom} \approx 10/v_{100}$ is much greater than $(1-f)(1-\bout ).$
Thus $L$, $M$, and $N$ are $> 0$ in all physical models.

Eq.\ (\ref{eq:harm}) is the equation for a harmonic oscillator,
$\ddot{y}+\gamma \, \dot{y}+\omega_0^2  \, y=0$.  If we define
$\Omega\equiv\sqrt{\omega_0^2-\gamma^2/4}$, then the criterion for
oscillation is $\Omega > 0$, which can be rewritten as 
\be 4 {\cal C}
N (1 - {\cal C}L) - (1 + {\cal C} M)^2 > 0.  
\ee 
In the limit in which
${\cal C}$ is very small or large this inequality is never satisfied, and
likewise at its local maximum ${\cal C} = (2 N-M)/(M^2 + 4NL).$ Thus we see
that for very a general class of models in which infall is modulated
by star formation, no oscillations are present in the
post-instantaneous recycling approximation.

As an additional check, we have solved eq.\ (\ref{eq:momback})
numerically over a broad range of parameters, blanketing the relevant
physical values.  These results confirm the analytical arguments
above, showing that oscillations never take place in models that
modulate infall as long as $\epsilon' \equiv \epsilon_{\rm
mom}-(1-f)(1-\bout)$ is positive.

\section{Conclusions}

While stellar feedback in galaxy formation has been the subject of
intense theoretical investigation, these studies have had little
impact on the direct interpretation of the observed properties of
elliptical galaxies.  Yet there are several observational clues that
point to its importance. Ferreras \& Silk (2000b) for example, found
that the color-magnitude relation of early-type cluster galaxies can
be easily understood in the context of a variable ejection efficiency
of supernova material, which scales inversely with galaxy mass.
Similarly, the large scatter in the faint end of the near-ultraviolet
minus optical color-magnitude relation in early-type cluster galaxies
hints at episodic star-formation in these objects, a natural
consequence of stellar feedback (Ferreras \& Silk 2000a). 

Motivated by these observational and theoretical pointers, we have
explored in this work a simple model of thermal feedback in elliptical
galaxies.  Our model consists of only gas and stars in a single zone
and is a natural extension of previous investigations (Tinsely 1980;
Ferreras \& Silk 2000b; Ferreras \& Silk 2001) which includes the
thermal state of the interstellar medium.  We account for heating by
supernova feedback and line emission cooling, and their impact on the
temperature and density of the ISM using a minimum of parameters, and
relate these quantities to the overall star formation rate using a
Schmidt law and a temperature-dependent efficiency.  This model is
applicable in the case in which the time for SNe to percolate through
the ISM is sufficiently long that spatial variations between regions
are relatively unimportant.  While this is the case for the choice of
parameters adopted in this paper, more detailed modeling is necessary
to study the interplay between thermalization and substructure
conclusively.

Our model leads to two important and independent results.  First the
interplay between infall and variable SNe ejection efficiency provides
a natural explanation of the optical and NIR color-magnitude relations
in elliptical galaxies both locally and at moderate redshift
($z\lesssim 1$).  Our model generates a mass-metallicity relation
$Z\propto M^{0.15-0.2}$, which accounts for the observed color range over
three magnitudes in luminosity. Furthermore, the rapid cooling times
for temperatures close to $10^4$~K result in short bursts of star
formation, generating stellar populations that are very similar to the
simple ones commonly used to describe ellipticals.  This implies no
change in the slope or the scatter of the optical and NIR
color-magnitude relation out to redshifts $z\lesssim 1$ as observed by
many authors (e.g.  Stanford et al.\ 1998; Van~Dokkum et al.\ 1998;
Van~Dokkum et al.\ 2000)

Secondly, we find that secondary peaks of late star-formation are
ubiquitous in smaller systems in which the fraction of the
SNe ejecta that escapes from the ISM, $\bout$, is significant.
In these cases the gas is heated to
temperatures $\propto (1-\bout)$, leading to secondary bursts of star
formation with a delay that scales as $\propto (1-\bout)^3$ due to the
$T^3$ scaling of the cooling times of high temperature gas.  As such
bursts occur within a Hubble time only in the smaller galaxies in
which $\bout$ is relatively large, this leads to a natural explanation
of the large scatter in the NUV-optical relation as observed in
clusters at moderate redshifts.  This can also account for the
observed population of post-starburst E+A galaxies that display a
spheroidal morphology (Ferreras \& Silk 2000c). While the current
claim for these post-starburst systems is the quenching of star
formation while falling through the hot intracluster medium (Poggianti et
al.\ 1999) our model shows that late starbursts may arise in
elliptical galaxies without resorting to environmental mechanisms.  In
fact, we have found that no such oscillations arise in a broad class
of theoretical models which study the interplay SNe and the further
accretion of gas. 

Given that the duration of these peaks of star formation is rather
short ($\sim 100$~Myr), the most promising observational approach is
to examine the NUV properties of spheroidal galaxies.  A comprehensive
study of the rest frame NUV$-$optical color-magnitude relation in
ellipticals may be able to quantify both the number of that undergo
late bursts and the mass fraction in young stars.  Furthermore, as the
thermal mechanism studied this paper is independent of environment,
comparisons of such colors between field, group, and cluster
populations can help to differentiate between this feedback and other
processes.  As the contribution in the NUV from evolved core
helium-burning stars becomes significant with age, studies of galaxies
at redshifts $z\gtrsim 0.3$ may provide the cleanest samples for such
comparisons.

At $z\gtrsim 1$, late bursts can be significant even in high-mass
galaxies, as shown in Figure~1.  This may result in the failure of
any search for high-redshift ellipticals that is based on
simple passively evolving models and does
not invoke complicated scenarios of assembly (e.g. Zepf 1997).  Our
model also underscores the bias intrinsic to observing luminosity
weighted quantities. While the mass and luminosity weighted ages of
higher mass ellipticals are quite similar, even the relatively small
bursts of late star formation that arise in smaller objects cause
large changes in their observed $V$ band luminosity-weighted ages.
This effect is even more severe in the NUV, highlighting the
importance accounting for the overall star-formation history when
interpreting observed stellar ages.

One of the remarkable results of our model is the presence of an
interstellar medium that has been heated by supernovae to temperatures
around $1$~keV or higher. This hot gas exists at very low densities,
with only a fraction leaving the galaxy, much in the same
way as the hot gas of the solar corona is kept gravitating around the
Sun.  A full 3D simulation would be needed in order to explore the
real conditions under which this ``fountain effect'' can occur.

Finally our prediction of a hot corona of SNe heated gas has natural
implications for the X-ray properties of elliptical galaxies.  By
approximating the overall X-ray luminosity of each object as due to
Bremsstrahlung emission from gas at the single temperature and density
given by our model, we derive a $L_X$ vs $L_B$ correlation that is in
good agreement with observed values.  Nevertheless our simple one-zone
model does not account for environmental effects and ignores what is
likely to be a significant contribution from unresolved X-ray binaries,
resulting in a scatter that is smaller than observed.

Galaxy formation is one of the richest and most complex processes in
all of astrophysics, forcing observers and theorists to approach it
from completely different viewpoints.  And although both theoretical
and observational progress has been clear and systematic, the rift
between theory and observation in the field remains one of the most
difficult to bridge.  In this work, we have attempted to draw the key
theoretical issue of feedback into the direct interpretation of the
observational properties of early-type galaxies.  While our approach
has been explorational, the widespread agreement of our simple model
with diverse observations suggests that thermal feedback processes are
likely to be essential to fully understanding the optical,
ultraviolet, and X-ray properties of early-type galaxies.

\acknowledgments 

IF is supported by a grant from the European Community under contract
HPMF-CT-1999-00109.  ES is supported by a National Science Foundation
MPS-DRF postdoctoral fellowship.  IF gratefully acknowledges the
hospitality of the astronomy department at the University of
California, Berkeley. We thank Ewan O'Sullivan for making available
the X-ray data of early-type galaxies used in this paper.

\end{document}